\pdfoutput=1
\newif\ifFull
\newif\ifShort
\Fulltrue
\ifFull
\else
\Shorttrue
\fi
\newif\ifAnon
\Anonfalse
\ifFull
\documentclass[11pt]{article}
\advance \topmargin by -\headheight
\advance \topmargin by -\headsep
\evensidemargin \oddsidemargin
\else
\documentclass{llncs}
\fi
\usepackage{color}
\usepackage{graphicx}
\usepackage{url}
\usepackage{epstopdf}
\usepackage{subfig}
\usepackage{verbatim}
\usepackage{array}
\usepackage{multirow}
\usepackage{bm}
\usepackage{sidecap}
\usepackage{times}

\ifFull
\makeatletter
\def\@begintheorem#1#2{\sl \trivlist \item[\hskip \labelsep{\bf #1\ #2:}]}
\def\@opargbegintheorem#1#2#3{\sl \trivlist
      \item[\hskip \labelsep{\bf #1\ #2\ #3:}]}
\makeatother
\fi

\newcommand{\E}{{\bf E}}

\ifFull
\newtheorem{theorem}{Theorem}

\fi

\newcommand{\eat}[1]{{}}

\begin{document}

\title{\LARGE Anonymous Card Shuffling and its \\ \LARGE  Applications to Parallel Mixnets}

\ifFull
\author{
{Michael T. Goodrich} \\
Dept.~of Computer Science \\ 
University of California, Irvine \\
goodrich@ics.uci.edu
\and
Michael Mitzenmacher \\
Dept.~of Computer Science \\ 
Harvard University \\
michaelm@eecs.harvard.edu
}
\fi

\date{}

\maketitle 


\pagestyle{plain}

\vspace{-0.20in}

\begin{abstract}
We study the question of how to shuffle $n$ cards when faced with an opponent
who knows the initial position of all the cards {\em and} can track every card
when permuted, {\em except} when one takes $K< n$ cards at 
a time and shuffles them in a private buffer ``behind your back,'' 
which we call {\em buffer shuffling}.
The problem arises naturally in the context of parallel
mixnet servers as well as other security applications.  
Our analysis is based on related analyses of load-balancing processes.
We include extensions to 
variations that involve corrupted servers and adversarially injected messages, 
which correspond to an opponent who can peek at some shuffles in the buffer
and who can mark some number of the cards.
In addition, our analysis makes novel use of a sum-of-squares metric
for anonymity,
which leads to improved performance bounds for parallel mixnets and 
can also be used to bound well-known existing anonymity measures.
\end{abstract}

\vspace{-0.25in}

\section{Introduction}
Suppose an honest player, Alice, is playing cards with a card shark, Bob,
who has a photographic memory and perfect vision.
Not trusting Bob to shuffle, Alice insists on shuffling the deck for each
hand they play. 
Unfortunately, Bob will only agree to this condition if he gets to 
scan through the deck of $n$ cards before she shuffles,
so that he sees each card
and its position in the deck, and if he also gets to watch her shuffle.
It isn't hard to realize
that, even though several well-known
card shuffling algorithms, like random 
riffle shuffling~\cite{MR841111}, 
top-to-random
shuffling~\cite{CambridgeJournals:1771344}, 
and Fisher-Yates shuffling~\cite{Knuth_1997}, 
are great at
placing cards in random order, they are
terrible at obscuring that order from someone like Bob who 
has memorized the initial ordering of the cards and is
watching Alice's every move.
Thus, these algorithms on their own are of little use to Alice.
What she needs is a way to shuffle that can 
place cards in
random order in a way that hides that order from Bob.
We refer to this as the \emph{anonymous shuffling} problem.
Our goal in this paper is to show that, as long as Alice has a 
private buffer where she can shuffle a subset of the cards, she 
can solve the anonymous shuffling problem. 

\begin{figure}[bt!]
\hspace*{-0.2in}
\begin{tabular}{cc}
\includegraphics[width=2.5in, trim=0.2in 0.4in 0.2in 0.4in, clip]{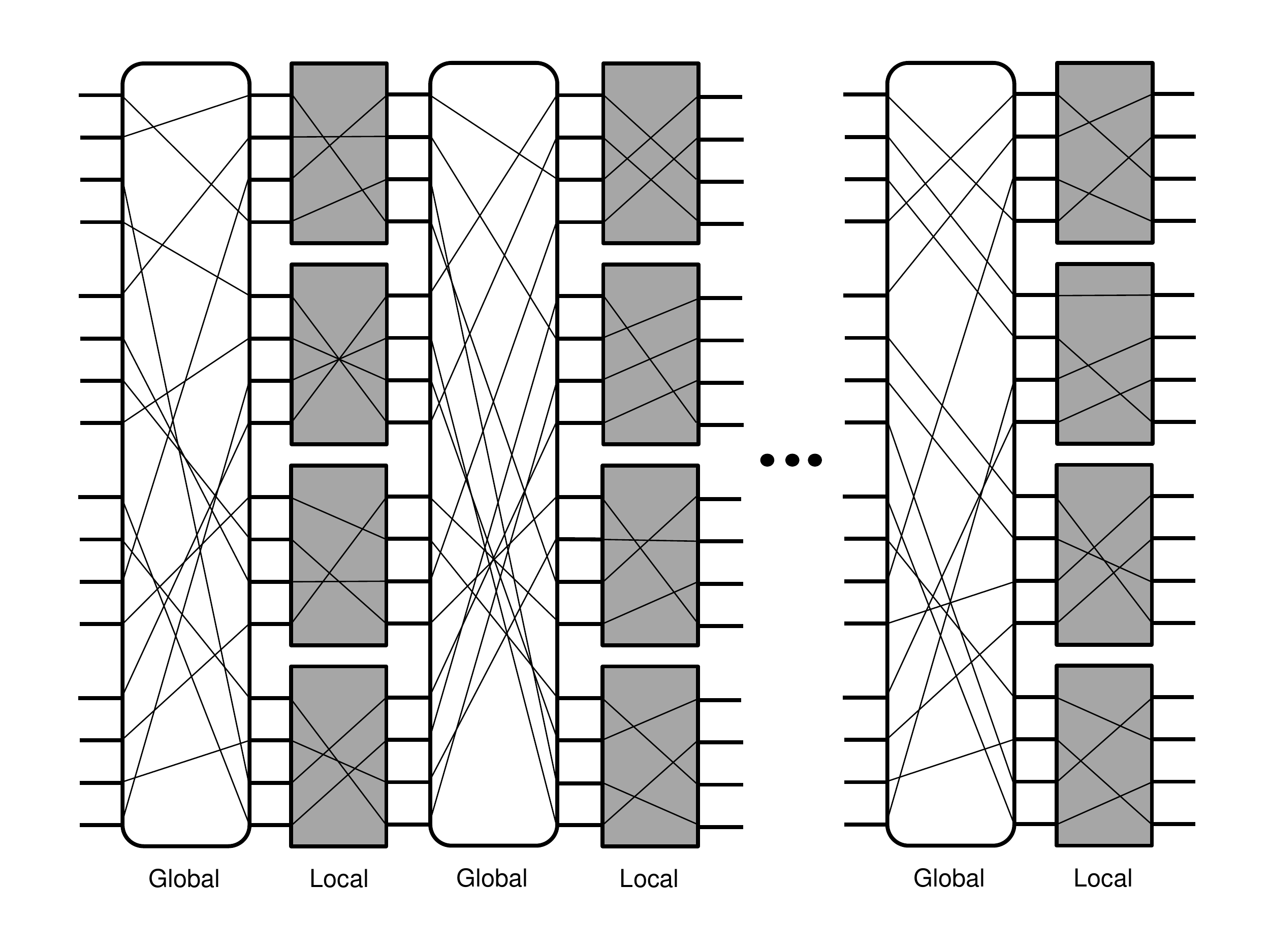} &
\includegraphics[width=2.5in, trim=0.2in 0.4in 0.2in 0.4in, clip]{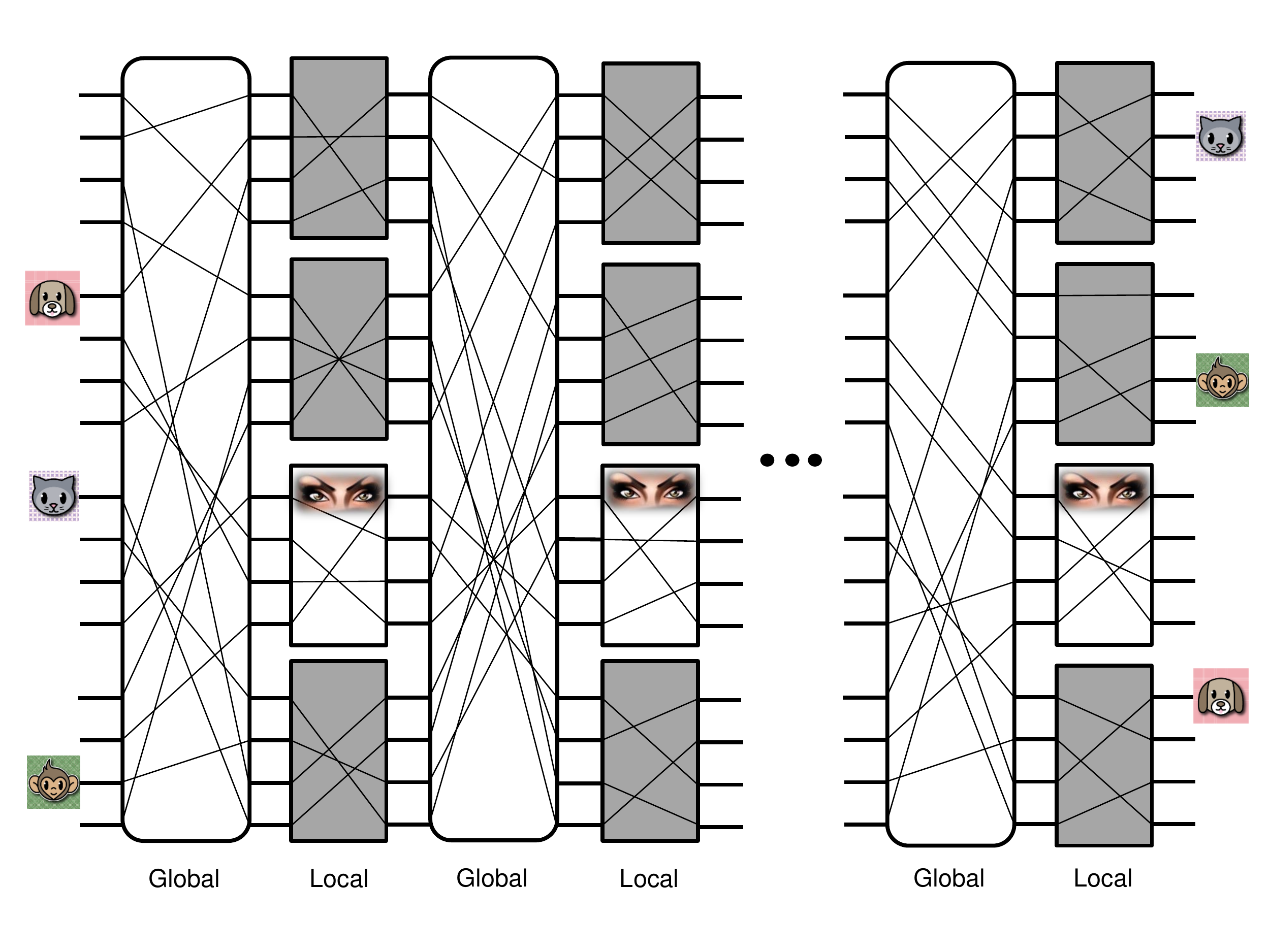} \\
(a) &
(b)
\end{tabular}
\vspace*{-10pt}
\caption{\label{fig:mixnet} (a) A parallel mixnet
with $n=16$ inputs and $M=4$ mix servers. Shaded boxes illustrate 
mix servers, whose internal permutations are hidden from the adversary.
The adversary is allowed to see the global permutation performed in each 
round.
(b) A corrupted parallel mixnet, where $s=3$ servers are not colluding with the
adversary, who has injected $f=3$ fake messages into the network.}
\vspace*{-12pt}
\end{figure}

Our main motivation for studying the anonymous shuffling problem in this paper 
comes from the problem of designing efficient parallel mixnets.
A \emph{parallel mix network}
(or \emph{mixnet}) is a distributed mechanism for connecting 
a set of $n$ inputs 
with a set of $n$ outputs in a way that hides the way the
inputs and outputs are connected.
This connection hiding is achieved by 
routing the $n$ inputs as messages through a set of $M$ 
\emph{mix servers} in a series of synchronized rounds.
In each round, the $n$ inputs are randomly assigned to servers so that each
server is assigned $K = n/M$ messages.
Then, each server randomly permutes the messages it receives and performs an
encryption operation so that it is computationally infeasible
for an eavesdropper watching the inputs and outputs of any (honest) server 
to determine which inputs are matched to the outputs.
The mixnet repeats this process for a specific number of rounds.
The goal of the adversary in this scenario is to determine 
(that is, link) one or more of the input messages with their corresponding
outputs, while the mixnet shuffles so
as to
reduce the linkability between inputs and outputs to an acceptably small
level.
(See Figure~\ref{fig:mixnet}a.)

Each of the servers is assumed to 
run the mixnet protocol correctly, 
which is enforced using cryptographic primitives 
and public sources of randomness 
(e.g., see~\cite{borisov-06,Golle:2004,kk-panm-05,Ren2010420,sp-smn-06}).
In some cases, we also allow for a \emph{corrupted} parallel mixnet,
where some number $s\ge 1$ of the servers behave properly,
but the remaining $M-s$ servers 
collude with the adversary so as to reveal how they are internally
permuting the messages they receive.
In addition, the adversary may also be allowed to inject
some number, $f < n$,
of fake messages that are marked in a way that allows the adversary to
determine their placement at any point in the process, including in the final output ordering.
(See Figure~\ref{fig:mixnet}b.)
In this paper, we are interested in studying a class of 
algorithms for anonymous shuffling, 
to show how the analysis of these algorithms can lead to improved
protocols for uncorrupted and corrupted parallel mixnets.

\subsection{Previous Related Work}

\ifFull
Arkin {\it et al.}~\cite{Arkin1999}
describe an attack on online poker based on exploiting a poor shuffling
algorithm, among other security weaknesses.
\fi

Chaum~\cite{Chaum:1981} introduced the concept of mix networks for achieving
anonymity in messaging, and this work has led to a host of other papers
on the topic (e.g., see~\cite{Ren2010420,sp-smn-06}).

Golle and Juels~\cite{Golle:2004} study the parallel mixing problem,
where mix servers process messages synchronously in parallel rounds,
and discuss the cryptographic primitives 
sufficient to support parallel mixnet functionality.
Their scheme has a total mixing time of $2n(M-s+1)/M$ and a number of 
parallel mixing rounds
that is $2(M-s+1)$,
assuming that $M^2$ divides $n$.
It achieves a degree of anonymity ``close'' to $n-f$, using a 
specialized anonymity measure, {\sf Anon}$_t$,
that they define (which we discuss in more detail 
in Section~\ref{sec:metrics}).
Note that if $s$ is, say, $M/2$, then their protocol requires as many rounds 
as the number of servers, which diminishes the potential benefits 
of a \emph{parallel} mixnet.
In particular, their approach 
uses sequences of round-robin permutations (cyclic shifts) rather than the standard
parallel mixing protocol described above and illustrated in
Figure~\ref{fig:mixnet}.
Even then, Borisov~\cite{borisov-06}
shows that their scheme can leak linkages between inputs and outputs 
(as can the standard parallel mixing protocol if the number of rounds
is too small) if the vast majority
of inputs are fake messages introduced by the adversary.
Thus, it is reasonable to place realistic limits on how large $f$ can be,
such as $f\le n/2$, and require that the number of parallel mixing rounds
is high enough to guarantee a high degree of anonymity.
Klonowski and Kutylowski~\cite{kk-panm-05} also study the anonymity of parallel
mixnets, characterizing it in terms of variation distance against the 
entire distribution for all messages. They consider
honest servers and the case of a single corrupted
server. 
They do not consider an adversary who can inject fake messages, however,
and they only treat the case when $M^2$ is much less than $n$.
Nevertheless, they show a lower-bound that says that full-distribution anonymity 
requires more than a constant number of rounds in the case of dishonest
servers.
Our approach avoids this lower bound, however, by focusing our measure of
anonymity on the obfuscation provided to individual messages,
as in other papers on parallel mixing  (e.g., see~\cite{Golle:2004,gmot-pos-12}), 
rather than to the entire distribution of messages, 
which allows us to achieve
anonymity in a contant number of rounds in some cases even in the presence of
dishonest servers. 
In addition, our focus on individual message anonymity is also motivated by 
the stated goal of mixnets, which is to provide anonymity to 
individuals sending messages through a parallel mixnet by combining their
messages together, rather than to provide anonymity for the entire set of users
as a group.



Goodrich, Mitzenmacher, Ohrimenko, and Tamassia~\cite{gmot-pos-12}
study a simple variant of the anonymous shuffling problem with no
corrupted servers or fake cards, addressing a problem similar to
parallel mixing in the context of oblivious storage. They show that
when the number of cards per server each round is $K = n^{1/c}$, then
$c+1$ rounds are sufficient to hide any specific initial card, so that
the adversary can guess its location with probability only $1/n
+o(1/n)$.  The current work provides a much more general and detailed
result, using much more robust techniques.

Our techniques are based on work in dynamic load balancing by Ghosh
and Muthukrishnan \cite{ghosh1994dynamic}.  In their setting, tasks
are balanced in a dynamic network by repeatedly choosing random 
matchings and balancing tasks across each edge.  Here, we extend
this work by choosing random subcollections of $K$ cards and balancing
weights, corresponding to probabilites of a specific card being one of
those $K$, among the $K$ cards via the shuffling.   

\subsection{Our Results}

We study the problem of analyzing 
parallel mixnets in terms of a buffer-based solution to the anonymous
shuffling problem,
assuming, as with other works on parallel 
mixnets~\cite{borisov-06,Golle:2004,gmot-pos-12,kk-panm-05}, that cryptographic
primitives exist to enforce re-encryption for each mix server, along
with public sources of randomness and permutation verification so that
servers must correctly follow the mixing protocol even if 
corrupted.

In the \emph{buffer shuffling} algorithm~\cite{gmot-pos-12,kk-panm-05},
Alice repeatedly performs a series of shuffling rounds, as in the parallel
mixnet paradigm.
That is, each round begins with Alice performing a 
random shuffle 
that places the 
cards in random order (albeit in a way that the adversary, Bob, can see).
Then she splits the ordered cards into $M$ piles, with each pile
getting $K = n/M$ cards.  
Finally, she
randomly shuffles each pile, using
a private buffer that Bob cannot see into.
Once she has completed her private shuffles, she stacks up her piles, which
become the
working deck for the next round. She repeats these rounds until she is
satisfied that the deck is sufficiently shuffled for the adversary.
Note that
during her shuffling,
Bob can see cards go in and out of her buffer, but he 
cannot normally see cards while they are in the buffer.
As we describe in more detail shortly, Alice's goal is to prevent Bob 
from being able to track a card;  that is, Bob should only be able to
guess the location of a card with probability $1/n+o(1/n)$, where
generally we take the $o(1/n)$ term to be $O(1/n^b)$ for some $b \geq 1$.  

To characterize the power of the adversary in the parallel mixnet
framework, we consider buffer shuffling in a context where,
for $M-s$ specific uses of Alice's buffer within each round, 
Bob is allowed to see how
the cards are shuffled inside it.  Likewise, we assume
he is allowed to mark $f<n$ of the cards in a way that lets him
determine their position in the deck at any time.  We
provide a novel analysis of this framework, and show how this analysis
can be used to design improved methods for designing parallel
mixnets.  For instance, we show that buffer shuffling achieves our goal
with $O(1)$ rounds even if the number of servers is relatively large and
that buffer shuffling can be performed in $O(\log n)$
rounds even for high degrees of compromise.
We summarize our results and how they compare with the previous
related work in Table~\ref{tbl:results}.

\begin{table}[hbt]
\vspace*{-5pt}
\begin{center}
\begin{tabular}{|c|c|c|c|c|c|}
\hline\hline
{\bf Solution} & {\bf Corruption} & {\bf Server}      & {\bf Allows for}
			& {\bf Allows for} & {\bf Rounds} \\
	        & {\bf Tolerance} & {\bf Restriction} & {\bf Corrupt Servers} 
			& {\bf Fake Messages} & \relax \\
\hline\hline
\rule[-6pt]{0pt}{18pt} 
GJ \cite{Golle:2004} & high & $c=2$ only & $\surd$ & $\surd$ & $O(M)$ \\
\hline
\rule[-6pt]{0pt}{18pt} 
KK \cite{kk-panm-05} sol'n 1 & none & $c=2$ only & $-$ & $-$ & $O(1)$ \\
\hline
\rule[-6pt]{0pt}{18pt} 
KK \cite{kk-panm-05} sol'n 2 & only one & $c=2$ only & $\surd$ & $-$ & $O(\log n)$ \\
\hline
\rule[-6pt]{0pt}{18pt} 
GMOT \cite{gmot-pos-12} & none & const. $c$ & $-$ & $-$ & $O(1)$ \\
\hline
\hline
\rule[-6pt]{0pt}{18pt} 
{\bf Our Theorem~\ref{thm:const}~} & medium 
	& const. $c$ & $\surd$ & $-$ & $O(1)$ \\
\hline
\rule[-6pt]{0pt}{18pt} 
{\bf Our Theorem~\ref{thm:log}~} & high 
	& const. $c$ & $\surd$ & $\surd$ & $O(\log n)$ \\
\hline\hline
\end{tabular}
\end{center}
\caption{\label{tbl:results} Summary of our results. We compare with the 
solutions of Golle and Juels~\cite{Golle:2004},
Klonowski and Kutylowski~\cite{kk-panm-05},
and Goodrich {\it et al.}~\cite{gmot-pos-12}. 
The server restriction column refers to the 
parameter $c$ in the inequality $K\ge n^{1/c}$.
The bounds of ``medium'' and ``high'' for corruption tolerance are made
more precise in the statement of the theorems.
Our theorems, as well as those of
Golle and Juels~\cite{Golle:2004} and Goodrich {\it et al.}~\cite{gmot-pos-12},
address anonymity from the perspective of individual messages;
Klonowski and Kutylowski~\cite{kk-panm-05} address anonymity for the entire
distribution of messages.
}
\vspace*{-8pt}
\end{table}

\section{Anonymity Measures}
\label{sec:metrics}

We can model the anonymous shuffling problem in terms of probability
distributions.
Without loss of generality, we can assume that the initial ordering of cards
is $[n]=(1,2,\ldots,n)$.
After Alice performs $t$ rounds of shuffling, let $w_i(t,c)$
denote the probability 
from the point of view of the adversary
that the card in position $i$ at time $t$ is the card
numbered $c$, and let $W(i,t)$ denote 
the distribution defined by these
probabilities (we may drop the $i$ and $t$ if they are clear from the context).
The ideal is for this probability to be $1/n$ for all $i$ and $c$,
which corresponds to the uniform distribution, $U$.

A natural way to measure anonymity is to use a distance metric to
determine how close the distribution $W$ is to $U$, for any particular 
card $i$ or in terms of a maximum taken over all the cards.
The goal is for this metric 
to converge to $0$ quickly as a function of $t$.

\paragraph{Maximum difference.}
The \emph{maximum-difference} metric, which is also known 
as the $L_\infty$ metric, specialized to measure the distance 
between $W$ and $U$, is 
\[
\alpha(t) = \max_{i,c} \left|w_i(t,c) - 1/n\right|.
\]
As mentioned above,
the goal is to minimize $\alpha(t)$, getting close to $0$ as quickly 
as possible.

Note that, 
in the case of buffer shuffling, the formula for $\alpha(t)$
can be simplified. 
In particular,
since Alice starts each round with a random
permutation,
$w_i(t,c)=w_i(t,1)$.
Thus, in our case, we can drop the $c$ and focus on 
$w_i(t)$, the probability that the $i$-th card is $1$.
In this case, we can simplify the definition as
\[
\alpha(t) = \max_{i} |w_i(t) - 1/n|.
\]

\paragraph{The ${\sf Anon}$ measure of Golle and Juels.}
In the context of parallel mixing,
Golle and Juels~\cite{Golle:2004} define
a measure for anonymity, which, 
using the above notation, would be
defined as follows:
\[
{\sf Anon}_t = \min_i \left( \max_c\, w_i(t,c) \right)^{-1} ,
\]
which they try to maximize.
Note that $\max_c\, w_i(t,c)\,\ge\, 1/n$
for all $i$, so,
to be consistent with the goals for other
anonymity measures, which are all based on minimizations, we
can use the following ${\sf Anon}'_t$ definition for an 
anonymity measure equivalent to that
of Golle and Juels:
\[
{\sf Anon}'_t = \max_i\left( \max_c\, w_i(t,c)\right) = \max_{i,c} w_i(t,c)
= ({\sf Anon}_t)^{-1}.
\]

The ${\sf Anon}'_t$
measure is not an actual distance metric, with respect to $U$, 
however,
since its smallest value is $1/n$, not $0$.
In addition, it is biased towards the knowledge gained by the adversary
for positive identifications and can downplay knowledge gained by ruling out 
possibilities.
To see this, note that, if we let $W^+$ denote all the $w_i(t,c)$'s that are 
at least $1/n$ and $W^-$ denote all the $w_i(t,c)$ values less than $1/n$,
then
\begin{eqnarray*}
\alpha(t) &=& \max\{ \max_{w_i(t,c)\in W^+} \{w_i(t,c) -1/n\}\, ,\, 
                  \max_{w_i(t,c)\in W^-} \{1/n - w_i(t,c)\}\} \\
         &=& \max\{ {\sf Anon}'_t -1/n\, ,\, 
                  \max_{w_i(t,c)\in W^-} \{ 1/n - w_i(t,c)\}\}.
\end{eqnarray*}
\ifFull
For example, suppose Alice shuffles the cards in a way that guarantees
that the first card is not the Ace of Spades but is otherwise as close
to uniform as possible.
Then, $\alpha(t)x=1/n$, since $w_1(t,1)=0$, 
and ${\sf Anon}'_t=1/(n-1)$ in this case.
Alternatively,
suppose Alice's shuffle is uniform except that it results in $w_1(t,1)=1/(n-1)$ 
and $w_1(t,c)=1/(n-1)-1/(n-1)^2$, for $c\not=1$.
In this case, 
$\alpha(t)=1/(n(n-1))$ while ${\sf Anon}'_t=1/(n-1)$, as in the other example.
The second example is much closer to uniform than 
the first and doesn't allow Bob to rule out
any specific card as being the first card, but the ${\sf Anon}'_t$ measure
(and, hence, the ${\sf Anon}_t$ measure) is the same in both cases.
\fi
Therefore, we prefer to use 
anonymity measures that are based on metrics 
and are unbiased measures
of the distance from $W$ to the uniform distribution, $U$.

\paragraph{Variation Distance.}
Li {\it et al.}~\cite{Li:2007}
introduce a notion of anonymity called
\emph{threshold closeness} or \emph{$t$-closeness}.
For categorical data, as in card shuffling and mixnets,
this metric amounts to the \emph{variation distance}
between the $W$-distribution defined by Alice's shuffling method
and the (desired) uniform distribution, $U$, where each card occurs with
probability $1/n$ (see also~\cite{kk-panm-05}).
In particular, this metric would be defined as follows for buffer shuffling:
\[
\beta(t) = \frac{1}{2} \sum_{i=1}^n | w_i(t) - 1/n |,
\]
which is the same as half the $L_1$ distance between the $W$-distribution
and the uniform distribution, $U$.
As with other distance metrics,
the goal is to minimize $\beta(t)$.

\paragraph{The sum-of-squares metric.}
\ifFull
In terms of measuring anonymity, an ideal metric is one that is sensitive to
outliers while still being easy to work with.
The reason that outlier sensitivity is useful is that focuses our attention
on the areas where the adversary, Bob, can gain the most advantage.
The above anonymity measures, related to the $L_1$
and $L_\infty$ metrics, have some sensitivity to outliers, but we would like
to use a metric that is more sensitive than these.
\fi

For this paper, we have chosen to focus on a metric for anonymity that is
derived from a simple measure that is well-known
for its sensitivity to outliers (which are undesirable in
the context of anonymity).
In this, the \emph{sum-of-squares}
metric, we take the sum of the squared differences
between the given distribution and our desired ideal.
In the context of buffer shuffling, this would be defined as follows:
\[
\Phi(t) = \sum_{i=1}^n ( w_i(t) - 1/n )^2,
\]
which can be further simplified as follows:
\begin{eqnarray*}
\Phi(t) &=& \sum_{i=1}^n ( w_i^2(t) - 2w_i(t)/n + 1/n^2) \\
 &=& \left(\sum_{i=1}^n w_i^2(t)\right) - 1/n.
\end{eqnarray*}
This amounts to the square of the $L_2$-distance between the $W$-distribution
and the uniform distribution, $U$.
The goal is to minimize $\Phi(t)$.

\ifFull
Incidentally,
this simplified form of the sum-of-squares metric, $\Phi(t)$, 
for the buffer shuffle,
doesn't take into account possible correlations between
pairs of items, but we show in this paper how to extend the sum-of-squares
metric to these contexts as well.
\fi

\paragraph{Relationships between anonymity measures.}
Another benefit of the $\Phi(t)$ metric is that it can 
be used to bound other metrics
and measures for anonymity, 
by well-known relationships among the $L_p$ norms.
For instance, we can derive upper bounds for other metrics 
(which we leave as exercises for the interested reader), such as
\[
\alpha(t) \le \Phi(t)^{1/2}
\mbox{\ \ \ \ and\ \ \ \ }
\beta(t) \le (n\Phi(t))^{1/2}/2.
\]
\ifFull
And we can also derive lower bounds for other metrics, such as
\[
\Phi(t)^{1/2}/2 \le \beta(t)
\mbox{\ \ \ \ and\ \ \ \ }
\Phi(t)^{1/2}/n^{1/2} \le \alpha(t).
\]
\fi
In addition, 
even though ${\sf Anon}'_t$ is not a metric,
we can derive the following bound for it, since 
${\sf Anon}'_t \le \alpha(t) + 1/n$:
\[
{\sf Anon}'_t \le \Phi(t)^{1/2}+1/n.
\]
So, for the remainder of this paper, we focus primarily on the $\Phi(t)$ metric.

\section{Algorithms and Analysis}
Our parallel mixing algorithm repeats the following steps:
\begin{enumerate}
\item Shuffle the cards, placing them according to a uniform permutation.
\item Under this ordering, divide the cards up into consecutive 
groups of $K=n/M$ cards.\footnote{As we also study, we could 
        alternatively assign each card uniformly at
	random to one of the $M=n/K$ piles, with each group getting
	$K=n/M$ cards in expectation.}
\item  For each group of $K$ cards, 
shuffle their cards randomly, hidden from the adversary.  
\end{enumerate}
We refer to each repetition of the above steps as a \emph{round}.
In the parallel mixnet setting, each group of $K$ cards would be shuffled at a different server.  

Let $w_i(t)$ be the probability that 
the $i$th card after $t$ rounds is the first card from time 0
from the point of view of the adversary.  
(We drop the dependence on $t$ where the meaning is clear.)
Initially, $w_1 = 1$, and $w_2 \ldots w_n$ are all 0.
Motivated by \cite{ghosh1994dynamic}, let $\Phi(t)$ be a potential function $\Phi(t) = (\sum w_i(t)^2) - \frac{1}{n}$,
based on the sum-of-squares metric,
and let $\Delta \Phi(t) = \Phi(t) - \Phi(t+1)$.  
(Again, we drop the explicit dependence on $t$ where suitable.)

Our first goal is to prove the following theorem.
\begin{theorem}
\label{thm:start}
A non-corrupted parallel mixnet, designed as described above, 
has $\E[\Phi(t)] \leq K^{-t}$.  In particular, such a mixnet,
with $K\ge n^{1/c}$, can mix messages 
in $t=bc$ rounds so that the expected sum-of-squares 
error, $\E[\Phi(t)]$,
between card-assignment probabilities and the uniform distribution is
at most $1/n^b$, for any fixed $b\ge 1$.
\end{theorem}
Before proving this theorem, we note some implications.
From Theorem~\ref{thm:start} and Markov's inequality, using $t=2bc$ rounds, we can bound the 
probability that $\Phi(t)>1/n^b$ to be at most $1/n^b$, for any fixed
$b\ge 1$.
So, taking $b=2$ implies $\alpha_t\le 1/n$ with probability $1-1/n^2$,
taking $b=3$ implies $\beta_t\le 1/n$ with probability $1-1/n^3$, and 
taking $b=2$ implies ${\sf Anon}_t'<(n-1)^{-1}$, with probability $1-1/n^2$,
which achieves the anonymity goal of
Golle and Juels~\cite{Golle:2004} (who only treat the case $c=2$).
Therefore, a constant number of rounds suffices for anonymously 
shuffling the inputs in a parallel mixnet, provided servers can internally
mix $K\ge n^{1/c}$ items, for some constant $c\ge 1$.

We now move to the proof.
Let $\Delta \Phi^*$ represent how the potential changes when a group of $K$ cards is shuffled during a round.  For clarity,
we examine the cases of $K=2$ and 3 before the general case. 
\begin{itemize}
\item For 2 cards with incoming weights $w_i$ and $w_j$ (outgoing weights are the average):
\begin{eqnarray*}
\Delta \Phi^* & = & w_i^2 + w_j^2 - 2((w_i+w_j)/2)^2 \\
            & = & (w_i - w_j)^2 /2.
\end{eqnarray*}
\item For 3 cards with incoming weights $w_i$, $w_j$, and $w_k$:
\begin{eqnarray*}
\Delta \Phi^* & = & w_i^2 + w_j^2 + w_k^2 - 3((w_i+w_j+w_k)/3)^2 \\
            & = & (w_i - w_j)^2/3 + (w_j - w_k)^2 /3 + (w_k - w_i)^2 /3 .
\end{eqnarray*}
\item For $K$ cards with weights $w_{i1}, w_{i2}, \ldots, w_{iK}$:
\begin{eqnarray*}
\Delta \Phi^* & = & \sum_{k = 1}^K w_{ik}^2 - K \left ( \frac{\sum_{k = 1}^K w_{ik}}{K} \right)^2 \\
            & = & \frac{1}{K}\sum_{1 \leq j < k \leq K}^K \left (w_{ij}-w_{ik}\right)^2.
\end{eqnarray*}
\end{itemize}

We now proceed to bound $\E[\Phi(t)]$ by making use of $\Delta \Phi$.    

\begin{eqnarray*}
\E[\Delta \Phi] & = & \frac{1}{K} \sum_{1 \leq i < j \leq n} \Pr \left( (i,j) \mbox{ are in the same set of } K \mbox{ cards} \right) (w_i - w_j)^2 \\
                & = & \frac{K-1}{K(n-1)} \sum_{i < j} (w_i-w_j)^2 \\
                & = & \frac{K-1}{2K(n-1)} \sum_{1 \leq i,j \leq n} (w_i-w_j)^2.
\end{eqnarray*}

Also,
\begin{eqnarray*}
\E[\Delta \Phi/\Phi] & = & \frac{K-1}{2K(n-1)} 
\frac{\sum_{i,j} \left( \left( w_i-1/n\right)-\left(w_j-1/n\right) \right)^2}{\sum_k \left(w_k - 1/n \right)^2}.
\end{eqnarray*}
Let $x_i = w_i - 1/n$ to get
\begin{eqnarray*}
\E[\Delta \Phi/\Phi] & = & \frac{K-1}{2K(n-1)} \frac {\sum_{i,j} (x_i- x_j)^2}  {\sum_k x_k^2}.
\end{eqnarray*}
Interestingly, when $K=n$, we should have mixing in one step, so in this case
$\E[\Delta \Phi/\Phi]$ should be 1.  Notice if that is the case, then perhaps surprisingly the above expression
is independent of the actual $x_i$ values, and then we have immediately:
\begin{eqnarray*}
\E[\Delta \Phi/\Phi] & = & \frac{n(K-1)}{K(n-1)}.
\end{eqnarray*}
We can in fact confirm this easily.  Since $\sum_k x_k = 0$, we have 
$$\sum_{i,j} (x_i- x_j)^2 = \sum_{i,j} (x_i- x_j)^2 + 2 \left(\sum_k x_k\right)^2 = 2n \sum_k x_k^2,$$
and cancellation gives the desired result.  

This analysis also gives us fast convergence to the uniform distribution
in the general case.  
Let $\gamma = \frac{n(K-1)}{K(n-1)}$, and note $\gamma \le 1$.  
In particular, 
$$1 -\gamma = \frac{n-K}{K(n-1)} < \frac{1}{K}.$$
Also note $\Phi(0) < 1$.  
So we have 
$$\E[\Phi(t+1)] = (1-\gamma) \E[\Phi(t)],$$ and a simple induction yields
$$\E[\Phi(t)] = (1-\gamma)^t \Phi(0) \leq K^{-t}.$$  
The rest of the theorem follows easily.


\subsection{Extensions to Mixnets with Corrupted Servers}

In the case of there being
corrupted servers, Bob will know the permutation for the cards
assigned to each such server.  In terms of the analysis, we can treat the
permutation for each corrupted server 
as the identity operation, since Bob can simply undo that
permutation.  
Let us suppose, then, that there are $M=n/K$ servers, so that each obtains $K$
cards in each round, and that $1\le s \leq n/K$ servers are uncorrupted.  
Following our previous analysis, we find
\begin{eqnarray*}
\E[\Delta \Phi] & = & \frac{1}{K} \sum_{1 \leq i < j \leq n} \Pr \left( (i,j) \mbox{ are in the same uncorrupted server} \right) (w_i - w_j)^2 \\
                & = & \frac{(K-1)}{K(n-1)} \frac{s}{n/K}\sum_{i < j} (w_i-w_j)^2 \\
                & = & \frac{s(K-1)}{2n(n-1)} \sum_{1 \leq i,j \leq n} (w_i-w_j)^2.
\end{eqnarray*}

Again, based on our previous analysis, we have
\begin{eqnarray*}
\E[\Delta \Phi/\Phi] & = & \frac{s(K-1)}{n-1}.
\end{eqnarray*}
Now let $\gamma' = \frac{s(K-1)}{(n-1)}$;
if, for example, $s = \epsilon \frac{n-1}{K-1}$ then 
$1- \gamma' = 1-\epsilon$.  
In that case,
$$\E[\Phi(t)] = (1-\epsilon)^t.$$  

\begin{theorem}
\label{thm:const}
A corrupted parallel mixnet, designed as described above, 
with $s\ge \epsilon(n-1)/(K-1)$ non-corrupted servers, 
for $\epsilon\ge 1/2$, can mix messages 
in $t=b\log n$ rounds so that 
the expected sum-of-squares 
error, $\E[\Phi(t)]$,
between card-assignment probabilities and the uniform distribution is
at most $1/n^b$, for any fixed $b\ge 1$.
Likewise, if there are at most $\frac{n^{-1/c}(n-1)}{K-1}-\frac{n-K}{K(K-1)}$ corrupted servers, 
with $K\ge n^{1/c}$ for some
constant $c\ge 1$, then in $t=bc$ rounds
it is also the case that
$\E[\Phi(t)]$ is at most $1/n^b$, for any fixed $b\ge 1$.
\end{theorem}

Thus, by Markov's inequality, using $t=2b\log n$ or $t=2bc$ rounds, 
depending on the number of uncorrupted servers, $s$, we can bound the 
probability that $\Phi(t)>1/n^b$ to itself be at most $1/n^b$, for any fixed
$b\ge 1$.


As an instructive specific example, suppose $K = M = \sqrt{n}$, and there are
a constant $z$ servers that are corrupted.  Then 
$$1 - \epsilon = 1 - (\sqrt{n}-z) \frac{K-1}{n-1} = 1 - \frac{\sqrt{n}-z}{\sqrt{n}+1}
= \frac{z+1}{\sqrt{n}+1}.$$
Hence, in this specific case, 
$$\E[\Phi(t)] = \left ( \frac{z+1}{\sqrt{n}+1} \right)^t < \frac{(z+1)^t}{n^{t/2}},$$
and for any constant $b$ after $4b$ rounds we have that 
$\Phi(t) \leq n^{-b}$ with probability $P(n^{-b})$. 

As our expressions become less clean in our remaining settings, 
we state a general
theorem which can be applied to these settings in a straightforward way: 
\begin{theorem}
\label{thm:log}
Given a parallel mixnet with corrupted servers or adversarially generated inputs,
let $\gamma =  \E[\Delta \Phi/\Phi]$ in that setting. 
Then in $t=b\log_{1/(1-\gamma)} n$ rounds 
the expected sum-of-squares 
error, $\E[\Phi(t)]$,
between card-assignment probabilities and the uniform distribution is
at most $1/n^b$, for any fixed $b\ge 1$.
In particular, if $\gamma \geq 1/2$, at most $b \log n$ rounds are required;  
if $\gamma \geq 1 - n^{-1/c}$, at most $bc$ rounds are required.  
\end{theorem}

\subsection{Extensions to Mixnets with Corrupted Inputs}

For the case of corrupted inputs, Bob will be able to track those cards throughout
the shuffle process.  In the shuffling setting, we can think of some
number of the cards as being marked---no matter what we do, Bob knows
the locations of those cards.
In terms of the analysis, we can treat this in the following way:
when we have a group of $K$ cards, 
it is as though we are shuffling only
$K' \leq K$ cards, where $K'$ is the number of unmarked 
cards in the collection of $K$
cards.  Let us suppose that $f \leq n-2$ cards are marked.  
Note that we may think of $w_i$ as being 0 for any cards in a marked position;  
alternatively, without loss of generality, let us calculate at each step as though 
$w_i$ is non-zero only for $i =1$ to $n-f$.  (Think of $w_i$ as being the appropriate
value for the $i$th unmarked card.)  Note that, for consistency, we must have
$$\Phi(t) = \left (\sum_{i=1}^{n-f} w_i(t)^2 \right ) - \frac{1}{n-f}.$$
Following our previous analysis, we find
\begin{eqnarray*}
\E[\Delta \Phi] & = & \!\sum_{K' = 2}^{K} \frac{1}{K'} 
                      \!\!\! \sum_{1 \leq i < j \leq n-f} 
                      \!\!\!\!\!\!\!\! 
       \Pr \left( \,\parbox{2in}{$(i,j)$ are in the same set of $K'$ out of $K$
                               unmarked cards} \,\right) (w_i - w_j)^2 \\
                & = & \sum_{K' = 2}^{K} \frac{{n-f \choose K'}{f \choose K-K'}}{{n \choose K}} \frac{K'-1}{K'(n-1)} \sum_{i < j} (w_i-w_j)^2 \\
                & = & \sum_{K' = 2}^{K} \frac{{K \choose K'}{n -K \choose n-f-K'}}{{n \choose f}} \frac{K'-1}{K'(n-1)} \sum_{i < j} (w_i-w_j)^2 \\
                & = & \frac{1}{2(n-1){n \choose f}} \left (\sum_{K' = 2}^{K} {K \choose K'}{n -K \choose n-f-K'} \frac{K'-1}{K'} \right) \!\sum_{1 \leq i,j \leq n-f} (w_i-w_j)^2.
\end{eqnarray*}

We can then compute
$\E[\Delta \Phi/\Phi]$ as 
\[
\frac{1}{2(n-1){n \choose f}} \left (\sum_{K' = 2}^{K} {K \choose K'}{n -K \choose n-f-K'} \frac{K'-1}{K'} \right) \frac {\sum_{1 \leq i,j \leq n-f} (x_i- x_j)^2}  {\sum_{1 \leq k \leq n-f} x_k^2}.
\]
Following the same computations as previously, we have
\begin{eqnarray*}
\E[\Delta \Phi/\Phi] & = & \frac{n-f}{(n-1){n \choose f}} \left (\sum_{K' = 2}^{K} {K \choose K'}{n -K \choose n-f-K'} \frac{K'-1}{K'} \right).
\end{eqnarray*}
Note the $n-f$ term in the numerator in place of an $n$.

Now let $\nu$ equal the right hand side above; then we have
$$\E[\Phi(t)] = (1-\nu)^t.$$  
In particular, it is clear that $\nu \leq \frac{n(K-1)}{K(n-1)}$, so the convergence of $\E[\Phi(t)]$ to 0 happens more slowly than in the case without corrupted inputs, as expected.
Nevertheless, we can still derive a theorem analogous to 
Theorem~\ref{thm:const} using 
Theorem~\ref{thm:log} and
the above characterization of $\E[\Phi(t)]$.
We omit a full restatement for space reasons.

\subsection{Extensions to Mixnets with Corrupted Servers and Inputs}

One nice aspect of our analysis is that combinations of corrupted servers and inputs are entirely straightforward.
In this setting, we have
\begin{eqnarray*}
\E[\Delta \Phi] & = & \sum_{K' = 2}^{K} \frac{1}{K'} \sum_{1 \leq i < j \leq n-f} 
\Pr \left(\, \parbox{1.9in}{$(i,j)$ are in the same set of $K'$ out of $K$
             unmarked cards at an uncorrupted server}\, \right) (w_i - w_j)^2 \\
                & = & \sum_{K' = 2}^{K} \frac{s}{n/K} \frac{{n-f \choose K'}{f \choose K-K'}}{{n \choose K}} \frac{K'-1}{K'(n-1)} \sum_{i < j} (w_i-w_j)^2 \\
                & = & \frac{sK}{n(n-1)} \sum_{K' = 2}^{K} \frac{{K \choose K'}{n -K \choose n-f-K'}}{{n \choose f}} \frac{K'-1}{K'} \sum_{i < j} (w_i-w_j)^2 \\
                & = & \frac{sK}{2n(n-1){n \choose f}} \left (\sum_{K' = 2}^{K} {K \choose K'}{n -K \choose n-f-K'} \frac{K'-1}{K'} \right) \sum_{1 \leq i,j \leq n-f} (w_i-w_j)^2.
\end{eqnarray*}

Hence
\begin{eqnarray*}
\E[\Delta \Phi/\Phi] & = & \frac{sK(n-f)}{n(n-1){n \choose f}} \left (\sum_{K' = 2}^{K} {K \choose K'}{n -K \choose n-f-K'} \frac{K'-1}{K'} \right).
\end{eqnarray*}

Given this bound,
we can then derive a theorem analogous to 
Theorem~\ref{thm:const}
for the case when mix servers can be corrupted and the adversary
can inject fake messages using Theorem~\ref{thm:log}.
We omit a full restatement for space reasons.

\ifFull
\subsection{Extensions to Mixnets with a Less Powerful Global Step}

Under our model, we can also consider a weaker global step between
rounds, where the cards must be split evenly into $n/K$ groups of
size $K$ for the $n/K$ servers.  Instead, let us assume simply that
each server obtains each card independently with probability $K/n$;
this requires less synchronization of the messages between rounds.
For example, previously we have assumed that the encryption step taken
by each server on each round is used to provide the random permutation
that maps messages to servers, and that a global step sorted the
re-encrypted messages in order for each server to obtain $n/K$
messages.  Assume instead that the encrypted messages are mapped to
values, which can be assumed to be uniformly distributed over their
range, and the range for such messages is divided into $n/K$ equally
sized subranges, one for each server.  Then the server for each
message for each round is determined without a complete sort (indeed,
at the end of reach round the behavior is like the first step of a
radix sort on random inputs).  Assuming $K$ is sufficiently large
(e.g., at least $c \log n$ for a suitable constant $c$) then Chernoff
bounds yield that each server will obtain at most $(1+\epsilon)K$ cards
in each round with high probability;  hence, the maximum shuffle size
can still be bounded.   

Extending the analysis above, we have that
$\E[\Delta \Phi]$
equals
\begin{eqnarray*}
\small
&  & \sum_{J=2}^n {n \choose J} \left ( \frac{K}{n} \right )^J \left ( 1- \frac{K}{n} \right )^J  \sum_{K' = 2}^{J} \frac{1}{K'} 
\!\!\!\sum_{1 \leq i < j \leq n-f} 
\!\!\!\!\!\!\Pr \left(\, \parbox{1in}{$(i,j)$ are in the same set of $K'$
            out of $J$ unmarked cards at an uncorrupted server}\, \right) (w_i - w_j)^2 \\
                & = & \sum_{J=2}^n {n \choose J} \left ( \frac{K}{n} \right )^J \left ( 1- \frac{K}{n} \right )^J \sum_{K' = 2}^{J} \frac{s}{n/J} \frac{{n-f \choose K'}{f \choose J-K'}}{{n \choose J}} \frac{K'-1}{K'(n-1)} \sum_{i < j} (w_i-w_j)^2 \\
                & = & \sum_{J=2}^n {n \choose J} \left ( \frac{K}{n} \right )^J \left ( 1- \frac{K}{n} \right )^J \frac{sJ}{n(n-1)}\sum_{K' = 2}^{J} \frac{{J \choose K'}{n -J \choose n-f-K'}}{{n \choose f}} \frac{K'-1}{K'} \sum_{i < j} (w_i-w_j)^2 \\
                & = & \sum_{J=2}^n {n \choose J} \left ( \frac{K}{n} \right )^J \left ( 1- \frac{K}{n} \right )^J \frac{sK}{2n(n-1){n \choose f}} \left (\sum_{K' = 2}^{K} {K \choose K'}{n -K \choose n-f-K'} \frac{K'-1}{K'} \right) \\
& & \ \ \ \ \ \cdot \sum_{1 \leq i,j \leq n-f} (w_i-w_j)^2.
\end{eqnarray*}

Generally, since for $K$ sufficiently large we will have concentration of the number of cards arounds its mean, the slowdown 
from the random distribution can be bounded by a small constant factor.

\fi

\section{Conclusion and Open Problems}
In this paper, we have provided a comprehensive analysis of buffer shuffling
and shown that this leads to improved algorithms for achieving
anonymity and unlinkability in parallel mixnets.
An interesting direction for future research could be to extend this analysis
to other topologies, including hypercubes and expander graphs.


\subsection*{Acknowledgments}
This research was supported in part by the National Science
Foundation under grants 
0713046, 0721491, 0724806, 0847968, 0915922, 0953071, 0964473, 
1011840, and 1012060,
and by the ONR under grant N00014-08-1-1015.
We would like to thank Justin Thaler for several helpful comments regarding
this paper.

{\raggedright
\ifFull
\bibliographystyle{abbrv}
\fi
\bibliography{refs,goodrich}
}

\clearpage
\begin{appendix}

\section{Additional Anonymity Measures}
There are several additional 
measures that Alice can use to determine when she is close to
placing her cards in random order in a fashion that
obscures that order from Bob.
As with the ${\sf Anon}_t$ measure, the ones we review here are also not
metrics.

\subsection{$k$-Anonymity}
One measure of obfuscation 
is \emph{$k$-anonymity}~\cite{SweeneyKanonymity02}. 
In the context of the anonymous shuffling problem,
we would say that Alice's shuffling algorithm achieves $k$-anonymity if,
from Bob's perspective,
every card in Alice's final output has at least $k$ 
input cards that have a possibility of being mapped to that card during the
shuffling.
This measure has been used in several applications in computer security and
privacy, including
mixnets~\cite{Ahn:2003,Wang:2007}.
A well-known weakness of 
$k$-anonymity (e.g., see~\cite{Hopper:2006,Li:2007,Machanavajjhala:2007}),
unfortunately, is that doesn't take probabilities into consideration.
So, for example,
if Bob has a 90\% certainty about the identity of the top card after Alice's
shuffle, with all the other cards sharing the remaining
10\%, then we would say that the top
card's identity had achieved $n$-anonymity (i.e., the maximum possible), 
even though Bob can be confident about which card is on top.
Thus, we feel that $k$-anonymity is insufficient to use as a measure of
obscurity for the anonymous shuffling problem.

\subsection{Relative entropy}
Relative entropy measures, in bits, the amount of information that exists
between two probability measures.
For probability distributions $P$ and $Q$, it is defined as
\[
D(P||Q) = \sum_i P_i \log_2 \frac{P_i}{Q_i} .
\]
In the context of buffer shuffling, and the information present
in the distribution, $W$, for a single card, 
compared to the uniform distribution, $U$, this amounts to
\begin{eqnarray*}
D(W||U) &=& \sum_i w_i(t) \log_2 \frac{w_i(t)}{1/n} \\
        &=& \log_2 n\ -\ \left|\sum_i w_i(t) \log_2 w_i(t)\right| .
\end{eqnarray*}
The goal for Alice would be to minimize the relative entropy, $D(W||U)$.
While relative entropy is more sensitive to outliers than the 
distance notations, $\alpha_t$ and $\beta_t$, related to the $L_\infty$ and
$L_1$ metrics, and it
measures information leakage in bits, which is a useful quantity, it is not
always easy to work with.
Moreover, it is not actually a metric, since it doesn't satisfy
the symmetric property.

\end{appendix}

\end{document}